\documentclass[12pt]{article}
\usepackage{graphicx}
\title{\bf Radial Velocity Variations in K Giants: Planets or Pulsations?}
\author{Saskia Hekker\thanks{saskia@strw.leidenuniv.nl},
Sabine Reffert and Andreas Quirrenbach \\
\vspace{1cm}\\
\normalsize Leiden Observatory, Netherlands}
\date{\mbox{}}
\begin{document}
\maketitle
\pagestyle{empty}
%
%
\def\bull{\vrule height .9ex width .8ex depth -.1ex}
\makeatletter
\def\ps@plain{\let\@mkboth\gobbletwo
\def\@oddhead{}\def\@oddfoot{\hfil\tiny\bull\quad
``Science Case for Next Generation Optical/Infrared Interferometric Facilities 
(the post VLTI era)'';
37$^{\mbox{\rm th}}$ Li\`ege\ Int.\ Astroph.\ Coll., 2004\quad\bull}%
\def\@evenhead{}\let\@evenfoot\@oddfoot}
\makeatother
%
%
\def\beginrefer{\section*{References}%
\begin{quotation}\mbox{}\par}
\def\refer#1\par{{\setlength{\parindent}{-\leftmargin}\indent#1\par}}
\def\endrefer{\end{quotation}}
%
%
{\noindent\small{\bf Abstract:} 
Radial velocity observations of K giants have revealed periodic variations for some of the stars. The periods range from approximately 180 to 1000 or more days with velocity amplitudes ranging from approximately 50 m/s to 500 m/s.  Except for Iota Draconis (Frink et al.\ 2002), all variations are nearly sinusoidal. Companions with zero orbital eccentricity as well as pulsations may cause these radial velocity variations. Additional high resolution spectra were taken in order to reveal which of these mechanisms is at work. No conclusions can be drawn from this work yet, although it seems likely that at least for some stars pulsations are the best explanation. A theoretical study is performed to investigate what kind of pulsations could cause these long period radial velocity variations. 
%
%
\section{Introduction}
Nowadays, techniques to perform radial velocity observations are well known and very accurate, up to a few m/s accuracy (see, e.g., Marcy \& Butler 2000, Queloz et al.\ 2001). Most extrasolar planets known so far have been discovered with this technique. Although most extrasolar planets are found around main sequence stars, a few have been discovered around giants (e.g.\ Frink et al.\ 2002, Setiawan et al.\ 2003 \& 2005, Sato et al.\ 2003).
For non-sinusoidal periodic variations of the radial velocity, a companion is a plausible explanation, but if the periodic variations are sinusoidal, these might as well be caused by pulsations.
Setiawan et al.\ (2003 \& 2005) calculated bisector velocity spans for their targets to see whether the spectral line shapes change or not, while Sato et al.\ (2003) disregarded pulsations for theoretical reasons. 

The present work reports on periodic radial velocity variations observed in K giants and provides a first attempt to reveal the nature of these periodic variations. 

\section{Radial Velocity Observations}\label{rvobs}
The radial velocity survey is performed at Lick Observatory, California USA, using the 60 cm Coud\'e Auxiliary Telescope (CAT) with the Hamilton spectrograph (R=60\,000). To obtain radial velocities with an accuracy of a few m/s an iodine cell is placed in the light path (Butler et al.\ 1996). The K giants are selected to be brighter than $V=6$~mag, presumably single and with masses ranging from approximately 1 to 3 solar masses. The survey started in June 1999 and a lot of stars have been observed more than 50 times by now.

So far, for one star a companion has been announced (Iota Draconis by Frink et al.\ 2002), but for a total of about 35 stars a periodic radial velocity variation is found. While the radial velocity variation for Iota Draconis is highly non-sinusoidal, indicating a high orbital eccentricity for the companion, for all other stars the radial velocity variation is nearly sinusoidal. This radial velocity variation may be caused by companions with zero orbital eccentricity as well as by pulsations.

\begin{figure}
  \begin{minipage}{8cm}
  \centering
  \includegraphics[width=6.5cm]{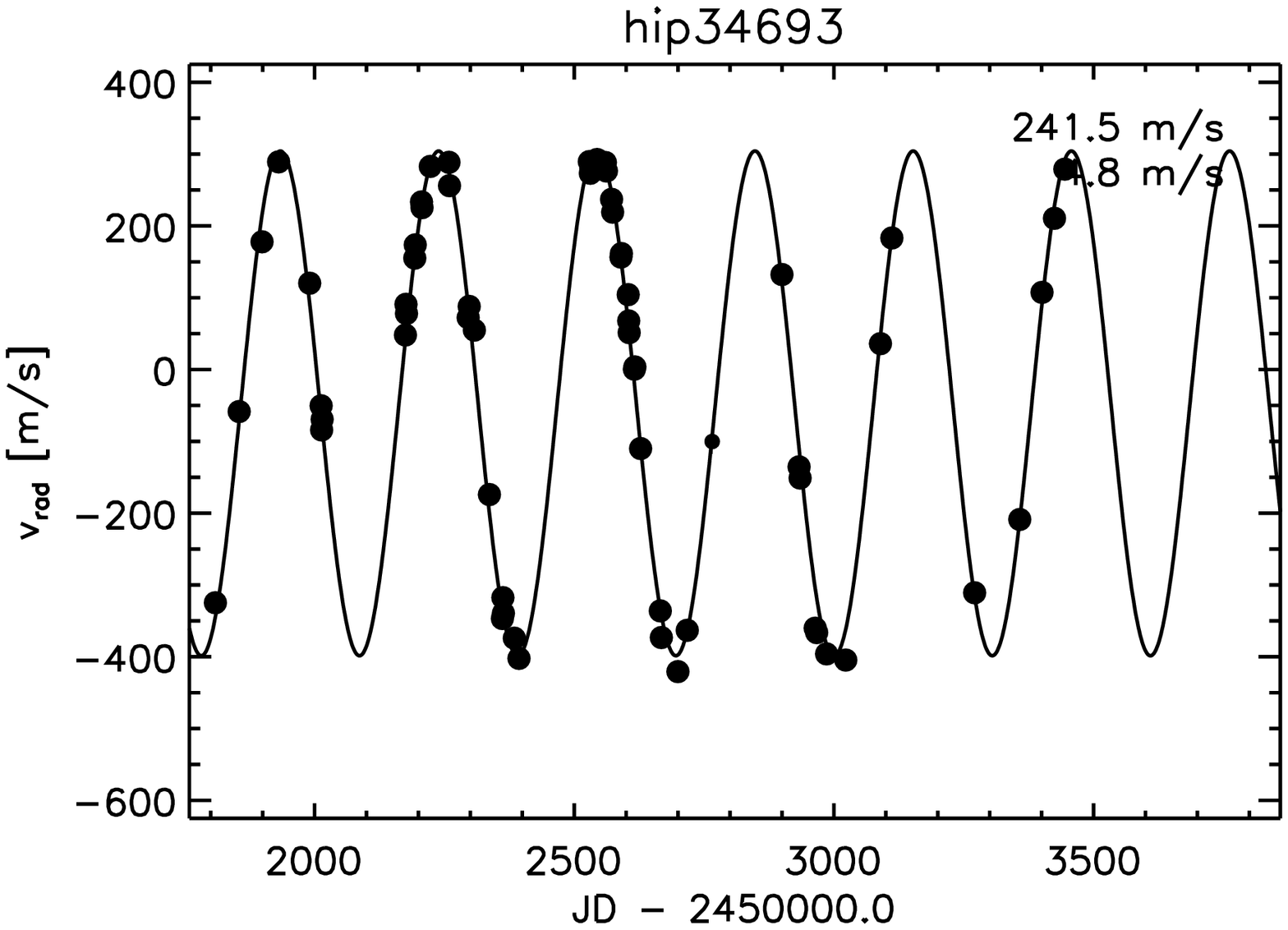}
  \caption{Radial velocity variation for hip34693 as a function of Julian Date. A Keplerian fit is plotted through the data: scatter around the fit is small.}
  \label{RVgood}
  \end{minipage}
  \hfill
  \begin{minipage}{8cm}
  \centering
  \includegraphics[width=6.5cm]{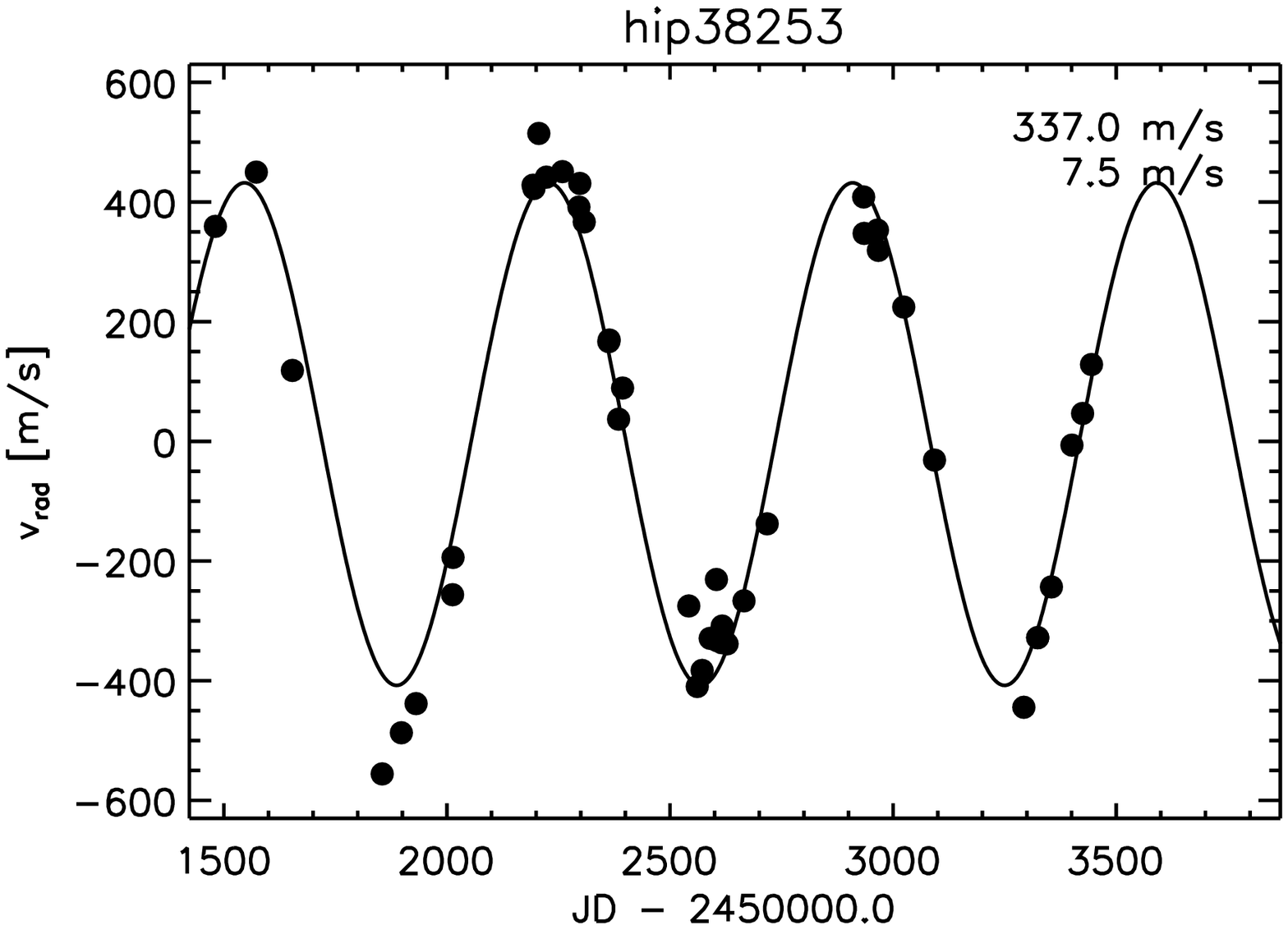}
  \caption{Radial velocity variation for hip38253 as a function of Julian Date. A Keplerian fit is plotted through the data: scatter around the fit is larger.}
  \label{RV}
  \end{minipage}
  \end{figure}

For each star a Keplerian orbit is fitted to the data. For four of the stars, the scatter around the plot is sufficiently small compared to the radial velocity amplitude while for all other stars the scatter around the fit is larger. In Figures~\ref{RVgood} and \ref{RV}, examples of both situations are shown. To see whether these four stars are really different from the other ones, the rms of the radial velocity is calculated as a function of colour. The stars with small scatter and Iota Draconis appear to have a much higher rms than other stars in the sample with the same colour.  Furthermore a closer look at the period distribution reveals that the stars with the best Keplerian fits have the shortest periods found.

Due to the better representation of the data by a Keplerian fit, the four stars with only small scatter around the Keplerian fit are more likely to host a companion than the other stars in the sample. An examination of the spectral line shapes may be able to discriminate between companions and pulsations.

\section{Line Bisectors} \label{bis}
To distinguish between a Doppler variation in case of a companion and changes in the spectral line shape in case of pulsations, high resolution spectra are needed. Therefore observations have been taken with the SARG high resolution echelle spectrograph on the Telescope Nazionale Galileo at La Palma (R=164\,000) and with the CES on the ESO 3.6 m telescope at La Silla (R=200\,000). An unblended line is selected in these spectra to perform the analysis. The bisector velocity span (BVS) and bisector velocity displacement (BVD) are calculated. 

A line bisector is defined as the difference between the center of a spectral line and the midpoint between the blue and red wing of the spectral line at a certain intensity. In case of the BVS, the bisectors at two different intensities (one in the upper half of the spectral line and one in the lower half) are subtracted, while for the BVD the bisectors at three different intensities are averaged (Povich et al.\ 2001). 

In case of a companion, both BVD and BVS remain constant over time, while in case of pulsations these parameters will change in time with the same period as the radial velocity variation. In Figure~\ref{bisv1}, the BVD for hip114855, one of the stars with small scatter around its Keplerian fit is shown as a function of its radial velocity variation. The BVD seems to be constant, which would indicate a companion. Figure~\ref{bisv2} shows the same plot for hip38253, which does not have such a good Keplerian fit. In this case the BVD changes as a function of the radial velocity variation, which would indicate pulsations. The data shown has been obtained with the CES.

 \begin{figure}
  \begin{minipage}{8cm}
  \centering
  \includegraphics[width=6.5cm]{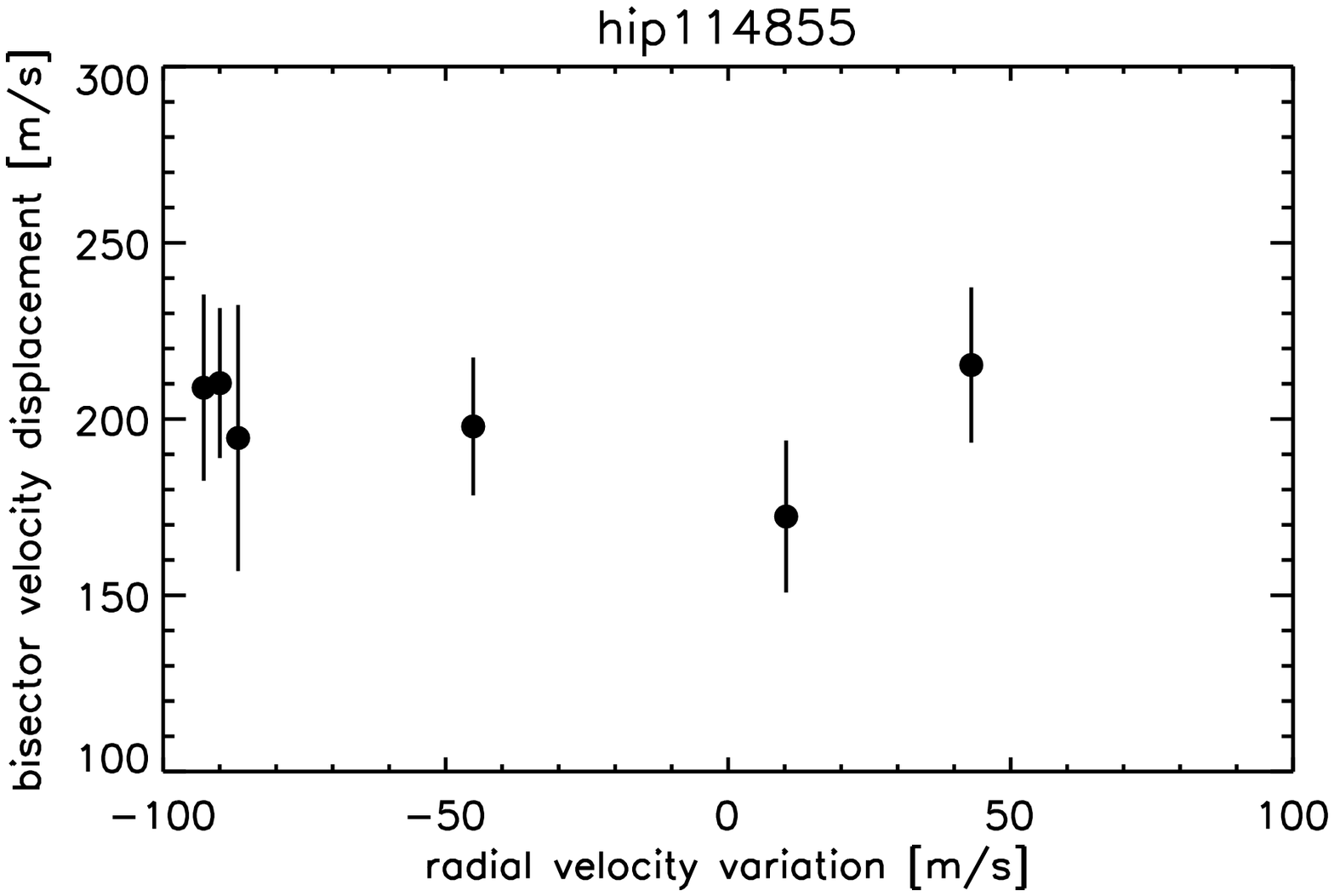}
  \caption{BVD for hip114855 as a function of the radial velocity variation.}
  \label{bisv1}
  \end{minipage}
  \hfill
  \begin{minipage}{8cm}
  \centering
  \includegraphics[width=6.5cm]{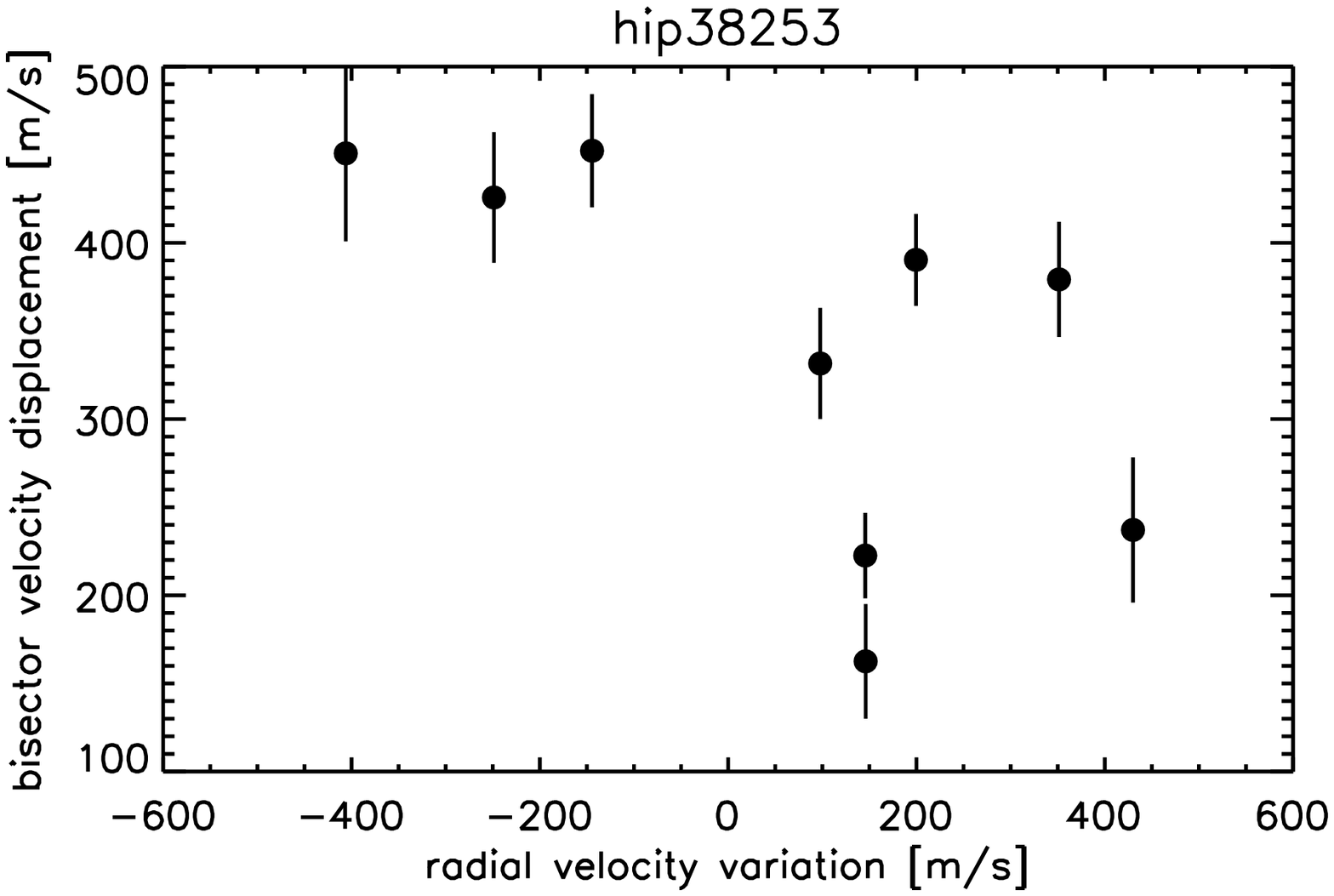}
  \caption{BVD for hip38253 as a function of the radial velocity variation.}
  \label{bisv2}
  \end{minipage}
  \end{figure}

As these are preliminary results, more observations and improvements in the data reduction have to be made in order to distinguish between companions and pulsations with more confidence. At this point, the BVD of some stars does hint to pulsations and therefore a theoretical idea of the kind of pulsations that might be able to produce these observed radial velocity variations is presented.

\section{Pulsations}\label{puls}
The pulsations most commonly observed in stars are the p-modes which have periods shorter than the fundamental period and cause a large vertical amplitude. The second type of pulsations are the g-modes, which have large horizontal amplitudes and periods longer than the fundamental period.

The fundamental periods of the K  giants in the sample range from approximately 10 hours to a few days. As the observed radial velocity variations have periods larger than 100 days this means that the radial velocity variations can, in case of pulsations, only be due to g-mode pulsations. Furthermore, the ratio between the horizontal and vertical velocity amplitude ($K$) which can be calculated from (Aerts et al.\ 1992):
\begin{equation}
K =\frac{horizontal~velocity~amplitude}{vertical~velocity~amplitude}=\frac{GM}{\omega^{2}R^{3}}
\label{K}
\end{equation}
 ($G$ is the gravitational constant, $M$ the mass of the star, $R$ the radius and $\omega$ is the frequency ($2\pi$ /period)), is larger than $10^{3}$. This implies that the horizontal velocity, i.e.\ the velocity tangential to the surface of the star, is much larger than the vertical velocity.  As this tangential velocity also contains the rotational velocity of the star, it is not justified to neglect the rotational velocity and the centrifugal and Coriolis force should be taken into account.

The centrifugal force would modify the shape of the star, while the Coriolis force would introduce an extra contribution to the velocity field. The centrifugal force is only important near 0.54 times the fundamental frequency (Townsend 2003), which is not in the range of the frequencies of the observed radial velocity variations and therefore this force can be neglected.  The Coriolis force is important for all frequencies lower than the fundamental frequency. This is the case for all K giants considered and therefore this force can not be neglected.

At very low frequencies, the Coriolis force will cause equatorial guiding. This means that perturbations around the equator of the star are enhanced at the expense of those at higher latitudes, and eventually  pulsations persist only along the equator. Equatorial guiding is shown in Figure~\ref{EQ}.

A different type of waves that can only occur in rotating bodies are the Rossby waves. These waves occur due to conservation of momentum. If a fluid parcel moves up or down from the equator it will experience a change in angular momentum. This will lead to a wave running over the surface in retrogade direction. This is shown in figure~\ref{RW}. Rossby waves are observed on the Earth, but not yet in a star.
\begin{figure}
  \begin{minipage}{8cm}
  \centering
  \includegraphics[width=5.cm]{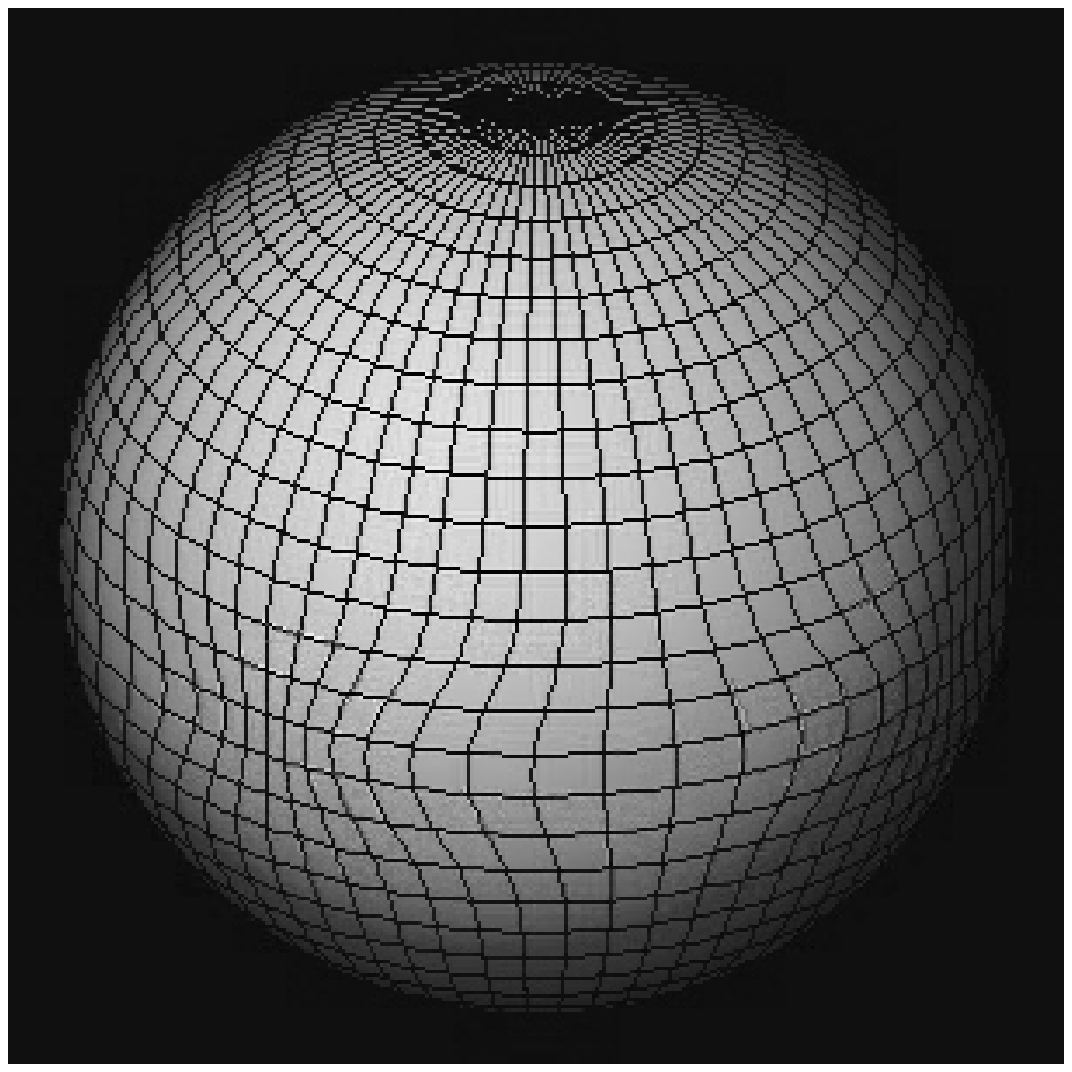}
  \caption{Rotating star with equatorial guided g-mode pulsations (Townsend 2003).}
  \label{EQ}
  \end{minipage}
  \hfill
  \begin{minipage}{8cm}
  \centering
  \includegraphics[width=5.cm]{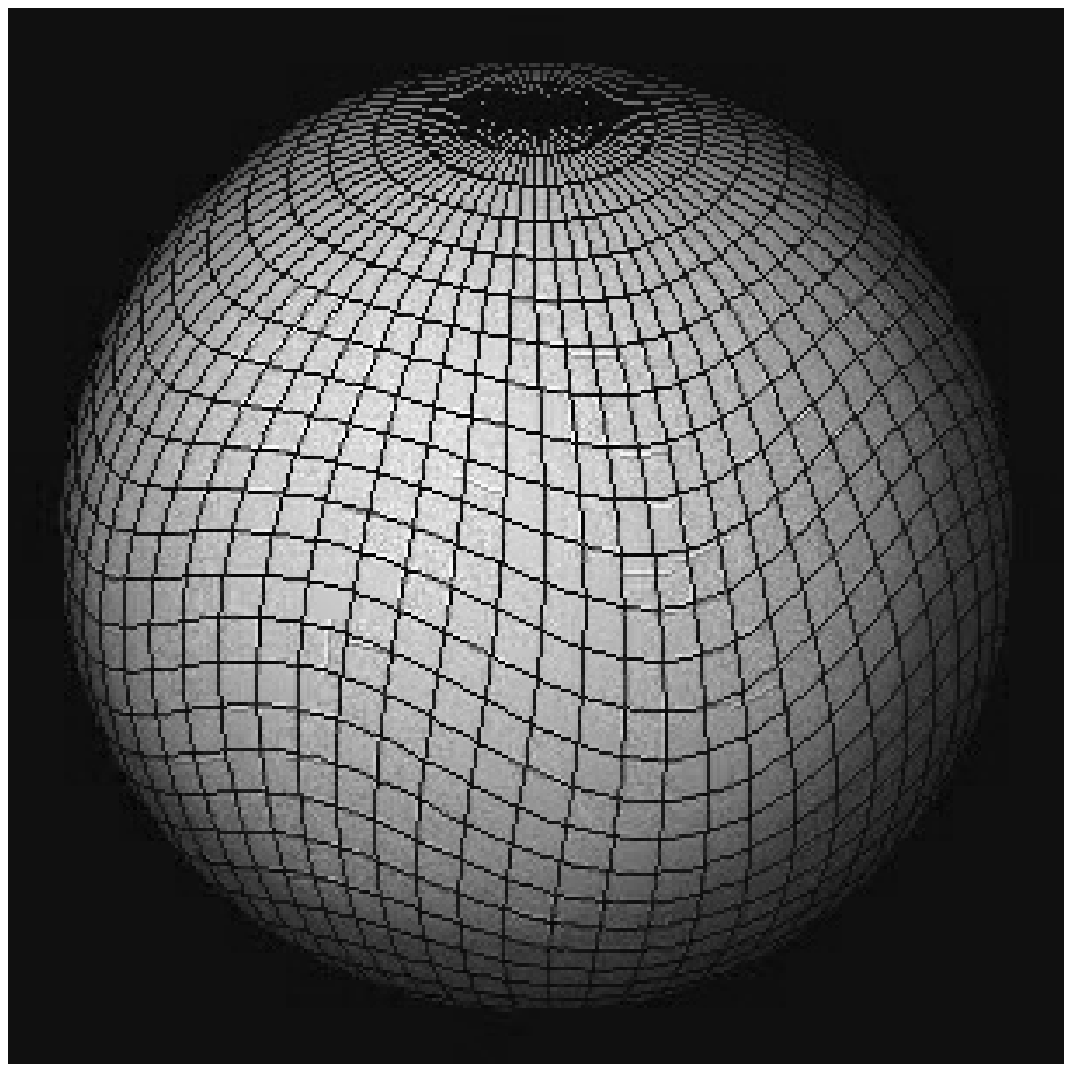}
  \caption{Rossby waves on an rotating star (Townsend 2003).}
  \label{RW}
  \end{minipage}
  \end{figure}

\section{Conclusions}
For all K giants described in this paper a clear periodic radial velocity variation is found and Keplerian orbits are fitted to the data of each star. As the radial velocity variations are nearly sinusoidal, a companion with zero orbital eccentricity as well as pulsations might cause these variations. To reveal which mechanism is at work an examination of a spectral line is performed using the BVD and BVS. The results are still preliminary and although for some stars the BVD and BVS hint to companions, for most stars they hint to pulsations. In case pulsations are present the moment method (Aerts et al.\ 1992), possibly with some extensions for rotation, might be useful to determine the order of the pulsation. 

Furthermore a theoretical idea is developed in order to know what kind of pulsations might cause these long period radial velocity variations. g mode pulsations with, to a certain extent, equatorial guiding seem to be most likely. Rossby waves might also be present, but little work has been done on the combination of these two phenomena.
%
%

%
%
 
\beginrefer
\refer Aerts C., De Pauw M., Waelkens C., 1992, A\&A 266, 294

\refer Butler R.P., Marcy G.W., Williams E., McCarthy C., Dosanjh P., Vogt S., 1996, PASP 108, 500

\refer Frink S., Mitchell D.S., Quirrenbach A., Fischer D.A., Marcy G.W., Butler R.P., 2002, ApJ 576, 478

\refer Marcy G.W., Butler R.P., 2000, PASP 112, 137

\refer Povich M.S., Giampapa M.S., Valenti J.A., Tilleman T., Barden S., Deming D., Livingston W.C., Pilachowski C., 2001, AJ 121, 1136

\refer Queloz D., Mayor M., 2001, Messenger 105, 1

\refer Sato B., Ando H., Kambe E., Takeda Y. et al., 2003, ApJ 597, 157

\refer Setiawan J., Hatzes A.P. Von der L\"uhe O., Pasquini L., Naef D., Da Silva L., Udry S., Queloz D., Girardi L., 2003, A\&A 398, 19

\refer Setiawan J., Rodmann J., Da Silva L., Hatzes A.P., Pasquini L., Von der L\"uhe O., De Medeiros R., D\"ollinger M.P., Girardi L., 2005, A\&A 437, 31

\refer Townsend R., 2003, IAU Symposium 215

\endrefer           
\end{document}